\documentclass [12pt,twoside]{article}           
\usepackage[left,modulo]{lineno}
\usepackage{setspace}

\countdef\arxiv=201

\ifnum\arxiv=0 \input cpreb.tex \fi

\ifnum\arxiv=1

\usepackage{epsfig,times,lscape}
\usepackage[usenames]{color}

    \ifnum\count102=1

    \fi

    \ifnum\count102=2
\fi

\newcommand{\nc}{\newcommand}

     \ifnum\count101=1
\nc{\qI}[1]{\section{{#1}}}
\nc{\qA}[1]{\subsection{{#1}}}
\nc{\qun}[1]{\subsubsection{{#1}}}
\nc{\qa}[1]{\paragraph{{#1}}}

\def\qpar{\vskip 2mm plus 0.2mm minus 0.2mm}
\def\qL{\hfill \break}
     \fi 

      \ifnum\count101=2
 \nc{\qI}[1]{\parindent=0mm \vskip 8mm 
{\centerline{\LARGE \color{red}#1}}\vskip 3mm}
\nc{\qA}[1]{\vskip 2.5mm \noindent 
{{\bf\large\color{blue}  #1}} \vskip 1mm \parindent=0mm}
 \nc{\qun}[1]{\vskip 1mm \noindent {\sl #1 }\quad }

\def\qL{\hfill \break}
\def\qpar{\vskip 2mm plus 0.2mm minus 0.2mm}

      \fi

\nc{\qfoot}[1]{\footnote{{#1}}}

      \ifnum\count101=1
\def\qbu{\hfill \par \hskip 6mm $ \bullet $ \hskip 2mm}
\def\qee#1{\hfill \par \hskip 6mm (#1) \hskip 2 mm}
      \fi
      \ifnum\count101=2
\def\qbu{\hfill \par \hskip 4mm $ \bullet $ \hskip 2mm}
\def\qee#1{\hfill \par \hskip 4mm (#1) \hskip 2 mm}
      \fi

\def\qparr{ \vskip 1.0mm plus 0.2mm minus 0.2mm \hangindent=10mm
\hangafter=1}

     \ifnum\count101=1 
 \def\qdec#1{\parindent=0mm\par {\leftskip=2cm {#1} \par}}
     \fi
     \ifnum\count101=2
  \def\qdec#1{\parindent=0mm \par {\leftskip=1cm {#1} \par}}
  
  \def\qcitb#1{\noindent \hbox to 102mm{\hfill \small #1} \vskip 1mm}
      \fi

 \def\qpages#1{\count102=0{\loop\advance\count102 by 1
 \null \vfill\eject \ifnum\count102<#1 \repeat}}

\def\qn#1{\eqno \hbox{(#1)}}

\def\qv{\vskip 0.1mm plus 0.05mm minus 0.05mm}

\def\qhv{\hskip 3mm}

\def\qhw{\hskip 1.5mm}
\def\qleg#1#2#3{\noindent {\bf \small #1\qhw}{\small #2\qhw}{\it \small #3}\qv }
\fi

\begin{document}

\thispagestyle{empty}

\centerline{\Large \bf \color{blue} Spatial and historical determinants}
\vskip 0.5cm 
\centerline{\Large \bf \color{blue} of}
\vskip 0.5cm 
\centerline{\Large \bf \color{blue} separatism and integration}
\vskip 0.5cm 
\centerline{\Large \bf \color{blue} 1. Qualitative analysis}
  
\vskip 1cm

\centerline{\large  Bertrand M. Roehner$^1$}
\qpar
\centerline{\large Institute for Theoretical and High Energy Physics,
University of Paris 6}

\vskip 1cm

{\bf Abstract}\quad Separatism is examined in a long-run
perspective. Accordingly, many political or economic factors which
may be crucial in dealing with short-term episodes can be safely
disregarded. Extending an approach pioneered by J. Jenkins, the
paper assesses the role of spatial and historical factors. It shows
that the means used to stage a separatist struggle are to a
notable extent borrowed from former historical episodes, an
analysis which supports and illustrates C. Tilly's thesis
of restricted repertoires of action. 
The
purpose of the present paper is to introduce the model, to make
it plausible and to demonstrate its potential usefulness by
examining a number of critical examples. A more systematic
analysis is carried out in a follow-up paper by using a data set
that includes about 40 cases of separatist struggles.

\vskip 1cm
\centerline{5 June 2017. Comments are welcome.}
\vskip 2cm

{{1: } Postal address: LPTHE, University Paris 6 (Pierre and Marie
Curie), 4 Place Jussieu, 75005 Paris   \qL
\phantom{1: } Email address: roehner@lpthe.jussieu.fr
\qL 
\phantom{1: } Phone: 33 1 44 27 39 16}

\vfill \eject
\def\qdecc#1{\parindent=0mm \par {\leftskip=8cm {#1} \par}}

\qdecc{\it ``The people are the land , and the land is the people.''}

\hfill Fijian proverb, cited in Robie (1989) 

\qpar

\qdecc{\it ``Language is the essence of human existence.''}

\hfill Maori proverb, cited in Fleras et al (1992)

\qpar

\qdecc{\it ``To speak the same language as one's neighbours express
solidarity with those neighbours; to speak a different language
from one's neighbours expresses social distance or even hostility''}

\hfill Leach (1954)

\large
\qpar

\qI{Introduction}

In the introduction of {\it ``Language and ethnic relations in
Canada''} Stanley Lieberson pointed out that ``the book stems from a
curiosity about why groups in contact maintain their distinctive
languages over the centuries in some countries, but elsewhere give
up their native tongues in a few generations''. This question is
at the heart of the present paper. What makes it perhaps even more
important in the late twentieth century 
is the following observation. Throughout the twentieth century the
number of sovereign countries has grown steadily, at first slowly and then,
after the Second World War, more and more rapidly (Figure 1).
  \begin{figure}[tb]
    \centerline{\psfig{width=16cm,figure=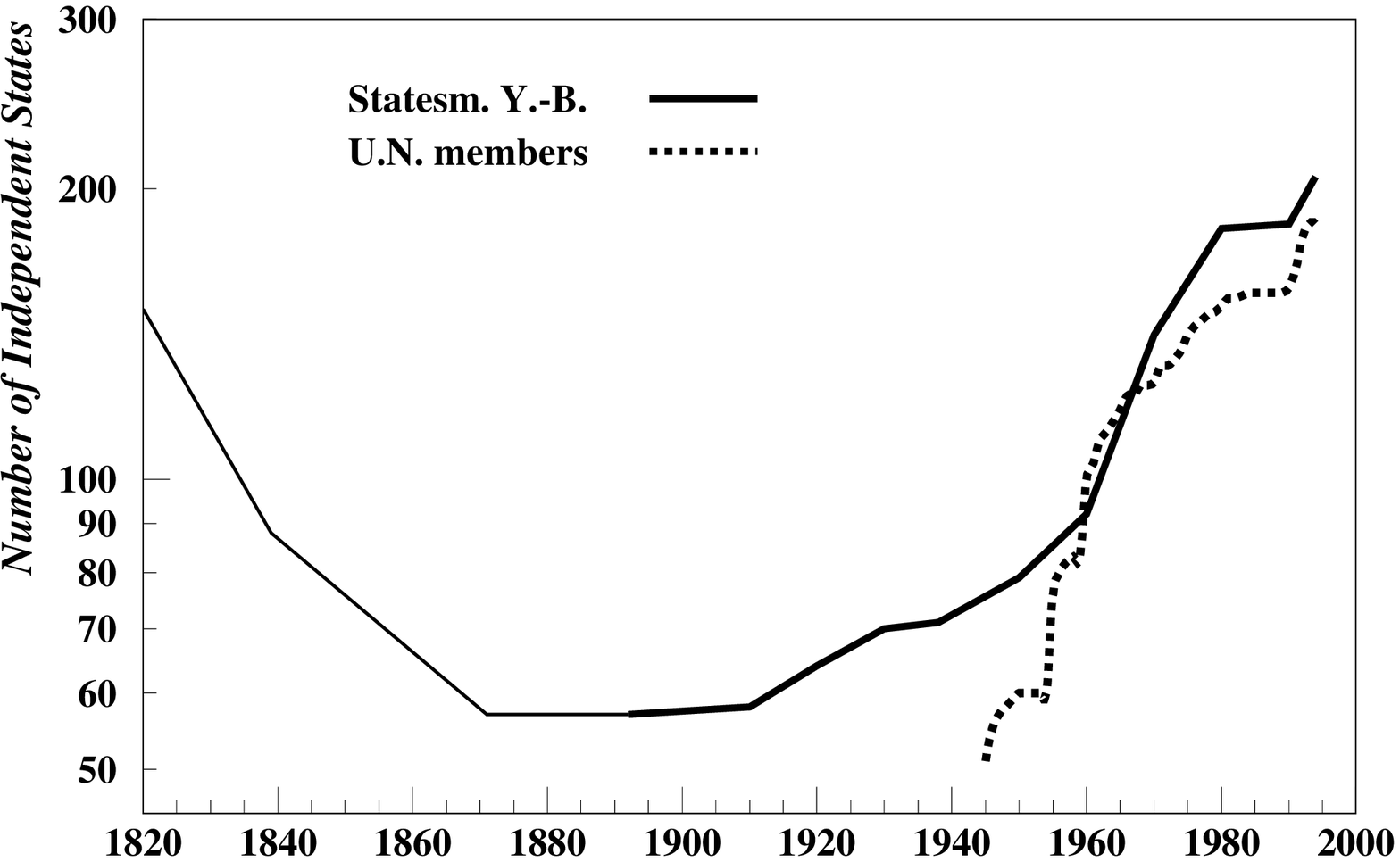}}
\qleg{Figure 1\qhv Evolution in the number of independent states.}
{The huge decrease
that occurred in the first half of the 
nineteenth century was mainly due to the
progressive unification of Germany. In 1803
there were about 500 sovereign or autonomous kingdoms,
bishoprics and other ecclesiastical territories; a first step was the
dissolution of the Holy Roman Empire in 1806; in 1833 there were still 31
German states. The independence of India in 1947 
occurred simultaneously with
the disappearance of about 140 local kingdoms and principalities;
it is unclear to what extent these entities could be considered
as really independent within British India which is why they were not
taken into account in the present chart.}
{{\it Sources:} {\it Statesman's Year-Book} (various years); 
Passant (1960); Quid
(1997); U.N. information leaflet about membership and admission dates.}    
 \end{figure}
The current 
trend sharply contrasts with the evolution during the nineteenth
century. Although the latter has been referred to as the 
century
of nationalities, a significant fall in the number of 
sovereign states occurred between 1850 and 1900, mainly as a 
result of the formation of the German Empire. The obvious question
then is whether in a long term perspective the current trend will
continue or whether a new coalescence cycle may set in%
\qfoot{Such a question is in fact beyond the scope of the present
paper; it requires a dynamic model whereas the one developed here
is a static, equilibrium model.}%
.
True, especially between 1950 and 1970, the decolonization process
accounted for the emergence of a great number of independent
states. Yet, a substantial number of new
sovereign states came into being through the disintegration of
former federations. While being the most obvious illustration, the
case%
\qfoot{Although the term ``disintegration'' is commonly used,
in fact it was rather an ordely separation in the sense that the
Republics which became independent (one can remember that Ukraine
had already been one of the founding members of the United Nations) had
the constitutional right to do so, whereas those which,
like the Chechen Republic, did not have that right remained in the
Federation. One must recognize that
the fact that the two types of entities were called
``Republic'' (albeit of different kinds) does not help 
to understand this distinction.}
of the USSR (Dec 1991) is not the only example; the
secession of Iceland from Denmark (1944), of Bangladesh 
from Pakistan (1971), of Somaliland
from Somalia (1991) and the scission of Czechoslovakia (January
1993) are other well known examples. Furthermore,  given the
number of separatist movements that are currently in progress, there
is a substantial probability  that the number of  sovereign states
will continue to increase in the near future.  Surprisingly enough,
while the trend in the political sphere has been toward a greater
degree of autonomy and federalism, in the economic sphere there has
been a general drift toward greater centralization. 
\qpar

There is already a vast literature on integration factors and on
separatist movements; a short review is given in section 2. In
general terms the two distinctive features of the present paper are
the following. 
\qee{1} The paper examines the long-term effect of spatial
factors. Only little attention has so far been given  to the
spatial determinants of separatism, a notable exception being the
work of Jenkins (1986) about Jura separatism in Switzerland. His
conclusion is worth quoting for it provides so to say the starting
point for the present study: ``This work concludes that two
geographical variables, the physical geography of the Bernese Jura
and distance were fundamental during the period of approximately
nineteen century following the birth of Christ in developing a
distinctive identity for the region which became the Bernese Jura
in 1815''. What
tends to hide the impact of spatial factors is the fact that in
short intervals (10 or 20 years) the role played by economic,
cultural and political factors is so conspicuous that it tends to
blur the slow but steady influence of geographical factors.
Furthermore, the latter do not usually 
 give rise to many spectacular events, a circumstance which makes
them even less visible. They operate at microhistorical rather than
at macrohistorical level.  Even a life-time may be too short a
period to observe any significant effect of geographical factors;
it is their steadiness which makes them important. Political and
economic factors may change, but geographical conditions remain
invariable. For instance, in the 19th century the French-speaking
part of Belgium was economically the leading province; 
unemployment was on average much higher in Flanders. By the end 
of the 20th century the situation is reversed. In contrast, an
island remains an island, a peninsula remains a  peninsula%
\qfoot{Man-made
transportation means may of course play a role; yet the
effect of such agencies is very slow in general. With a tunnel under
the English Channel, Britain is ``less'' an island than before.  
Howeveror, for the time being, ,
the fare is about the same whether one takes the tunnel or a
ferry.}%
. 
\qee{2} Our
approach is very much in line with the approaches of Connor (1972),
Lieberson (1970,1975) and Tilly (1986,1993). More
specifically, it owes much to Connor's emphasis on
 historical factors; it relies on
Lieberson's analysis of language and demographic
determinants; and at the  methodological level our strategy in
implementing events analysis follows rather closely the methods
pioneered by Tilly and Olzak (1992). 
\qee{3} Our primary objective
is  to confront our model with empirical quantitative evidence.
It is
chiefly for the purpose of testability that our model uses
only a small number of parameters.  In so doing we do not wish to 
deny that other causes are present. Many ethnic, cultural or
political factors are left aside, not because they are unimportant
but because it would be difficult to include them in a comparative
analysis based on quantitative data. While being probably
unacceptable in a short-term perspective, such a drastic selection
is less questionable in a long-run analysis.

\qA{Presentation of the model's main orientations}

There are two rather distinct part in this study. Our objectives in
the first part are (i) To delineate the scope of the model by
distinguishing those phenomena to which it applies and those  which
are outside its field of applicability (ii) To motivate our
selection of the model's parameters by a number of
arguments. In so doing, we frequently proceed by way of
illustrations and examples; our aim is to suggest rather than to
prove. Although these examples could possibly be omitted in a more
formal  presentation of the model, they are intended
to emphasize its empirical roots. Illustrations cannot
replace statistical tests however; the latter are provided in the
second part of the study where the model is confronted with a fairly
systematic body of empirical evidence. 
\qpar

The main purpose of this paper is to investigate the properties
and the implications of what could be called the blend of language
and homeland. The last term refers to a territorial base which
has been occupied for a ``long time'' by a people. The importance
of these elements is obvious: language is the medium of
communication and, especially in rural societies, there is a
strong connection between a people and its homeland . As a matter of
illustration let us consider the example of the United
States versus Switzerland. Both countries have a federal
organization; yet, in terms of integration they turn
out to be very different. The linguistic frontier of the two main
components of the Swiss confederation, the French- and the
German-speaking peoples, have remained practically unchanged for
centuries. As will be shown subsequently  the
linguistic frontier is also a very effective barrier to domestic
migration; even worse, opinions on major political issues such as
the European construction are largely determined by language
subdivisions.  In short, integration seems to have made little
progress. The situation is completely different in the United
States. In order to emphasize  how strongly many American national
subgroups tried their best to preserve their mother tongue, it is
perhaps of some interest to recall the following quotes.\qL  ``Few
of their children learn English. Of the six printing houses in the
province [Pennsylvania], two are entirely German, two half
German-half English. They have one German newspaper. The signs on
the street have inscriptions in both languages.''[In 1787 German
Americans represented about 9 percent of total population.] 
 
\hfill Letter of B. Franklin to a British Member of Parliament,
May 9, 1753; cited in Crawford (1992).

\qL
``In 1840 parents in Cincinnati demanded and obtained bilingual
education for their children. Not only in Cincinnati, but in many
communities in the latter half of the 19th century, education
using the native tongue of the non-English-speaking immigrants
florished.''

\hfill Thernstrom (1980,p.619)

\qL
``By 1914 the National German American Alliance had an impressive
membership of approximately two million people. At that time the
German American press included over 500 German-language
publications. In 1918 over 7,000 enemy aliens were arrested by
the Justice Department, many of them German Americans. Of the 200
German-language publications that survived World War I, only 24 
remained in 1976.''

\hfill Thernstrom (1980,p.685)

\qpar

These few facts remind us  that the attachment of the German
community to its language and to its ``Vaterland'' did not disappear
easily. At the same time it shows the strength of an
integration process that was able to overcome the nationalist
feelings not only of the Germans but of many other peoples who
were no less attached to their language and to their homeland.
Why did the melting-pot process work in the American case and not at
all in the Swiss case? Many specific reasons certainly can and have
been invoked to explain that particular case. In this paper we are
rather interested in {\it general factors}, that is to say causes
which can be shown to be at work in a large variety of similar
situations. In that spirit one may mention the following
differences: 
\qee{1} The linguistic border between American Germans
and the rest of the population was of much greater length than in
the case of Switzerland, the main reason being that the area
covered by American Germans was made up of several small domains; in
Switzerland on the contrary it formed (more or less) one domain%
\qfoot{In mathematical terms, the first domain would be referred to
as a multiply connected domain, while the second domain is said to
be simply connected.}%
.
\qee{2} The American German community
always represented a small share (less than 15 percent) of the
total American population; in contrast the Swiss French-speaking
community accounted for 20 to 30 percent of the Swiss population.
\qee{3} Finally one may mention an historical reason.  The
time interval between emigration and assimilation was to short for
American Germans to constitute a homeland in the United States. On
the contrary in the case of Switzerland there was plenty of time for
the cement between homeland and language to set hard.    \qpar

The above example suggests three factors of significance:
\qee{1} The length of the linguistic frontier (along with its
width, see below)
\qee{2} The relative proportion of the subgroup in the total
population. 
\qee{3} The degree of identification a people has established in
the course of history with its territorial base.  \qL
We posit that these factors account for a large number of
separatist movements in industrialized as well as in developing
countries. The influence of the mobilization parameter which has
been thoroughly studied, in particular by K. Deutsch, is represented
in our model by the width of the linguistic frontier. One should
note that the model does not apply to minorities characterized by
non-linguistic ethnic factors. Thus, for instance, the following
minorities fall outside the scope of the model: Mormons,
Hutterites, Jews and in general all religious minorities (the
question of the respective roles of religion and language is
discussed below), Blacks in the Unites States,
Burakumins in Japan, etc.

\qI{Methodological preparation}

\qA{Short survey of the literature}

Although there is a vast literature on national integration and
separatist movements, the studies focusing on the bond
between mother tongue and homeland are not so numerous. It
it is hardly possible in the framework of this paper
to give an account of the various models
that have been proposed: comprehensive and very readable reviews
are to be found in Connor (1994) and Premdas (1990) for instance.
In this paragraph we propose a classification of recent
contributions according to the kind and extent of the empirical
evidence with which the models have been confronted. 
 It is possible  to distinguish four categories in that respect.
 \qee{1} Purely theoretical models, i.e. models which have not yet
been confronted with any piece of empirical evidence. 
\qee{2} Models that have been confronted with qualitative
evidence. 
\qee{3} Models that have been confronted with a limited amount of
quantitative evidence. 
\qee{4} Models that have been confronted with extensive
statistical evidence. 
\qpar

Three contributions in the last category will be discussed
specifically because of their direct connection with the models
developed in this paper. \qL

{\bf Lieberson et al. (1975)} \quad This pioneering study examines
the determinants of mother tongue diversity. It analyzes the
evolution of language diversity in 35 states and over periods
ranging from a century to a few years, depending on data
availability. Various national characteristics are considered in
relation to changes in language diversity. Two factors, the
spatial isolation of language groups and official educational
policies turn out to have a significant influence on language
diversity. Spatial isolation was estimated through an index
proposed by Bell (1954). In a sense, given the reliance on indexes
and aggregated figures it can be said that the analysis by Lieberson
et al.  considered the problem in a macrosociological
perspective. In contrast, through its emphasis on basic mechanisms,
the present paper rather presents a microsociological view. 
\qL

{\bf Allardt (1979)} \quad This short book analyzes 46 linguistic
minorities in Western Europe, ranging from large subgroups such as
the Catalans (5.7 million) to small minorities such as the Faroe
islanders (46,000). As many as 18 variables are considered
describing the minorities in various aspects: demography,
education, welfare, politics, etc, but little attention is given
to the historical background.  Cross-correlations for all 
variables are systematically computed; most 
correlations turn out to be too small to be
significant; yet, more embarrassing is the fact that no clear
pattern seems to emerge from the few cross-correlations which are
high enough to be significant. 
\qL

{\bf Gurr (1993)} \quad Ted Gurr's analysis of communal
mobilization relies on an impressive and unique statistical data
base. Basic characteristics (population, ethnic classification,
etc) are provided for 227 communal groups. It covers
a the whole period from Second World War to present. The statistical
inquiry which we develop in the second part of this study in many
respects follows Gurr's approach. One marked difference is that
Gurr does not restrict his study to minorities with a definite
territorial base: religious minorities (Jews in Argentina for
instance) or ethnic minorities (foreign workers in Switzerland for
instance) are considered along with regional minorities. The task
of modeling such  a large class of minorities is probably even
more complex than the one we take up in this paper.

\qA{``Doable versus undoable questions''}
The distinction between so-called ``doable'' and ``undoable''
questions has been put forward by S. Lieberson (1985). We believe it
to be very important. For policy purposes one would want to
predict the outcome of separatist struggles that are currently in
progress or to identify  those that are likely to erupt in the
future. These, however, are very difficult problems. An analogy
proposed by S. Lieberson (1985, p.97) makes the matter very clear.
``Suppose I wish to find out the factors that lead a small number
of people to reach stardom in motion pictures. It might well be
possible to find certain characteristics of the individuals which
affect their chances, for instance personal attractiveness,
persistence, connections, even acting talent. Efforts to go beyond
the establishment of probability functions for each type of
characteristics are doomed to failure. Sure, every special case
could in a sense be `explained'; but the relevance of such ad
hoc, a posteriori reasoning is doubtful''.
 In the same way as the achievement of stardom,
the emergence or the success of a separatist struggle sequentially
depends  upon a large number of special circumstances. For instance
even a charismatic leader is unable to play any political role
unless the state has been compelled (usually by some external
pressure) to renounce to ruthless repression%
\qfoot{A spectacular example of such a relentless repression is
given by the suppression of the rebellion in East-Timor between
1976 and 1988.}%
.
Thus, trying to answer questions of interest for
policy purposes  would probably make us waste time and
energy on ``undoable'' questions. What is worse, it could lead us to
disregard, as being of little interest, precisely those questions
that may be settled conclusively. 
\qpar

For those who dislike analogies the above argument can be presented
in a different, more mathematical way. Basically,
separatist struggles are transient, non-equilibrium episodes.
Roughly  speaking one may say that there are two opposite forces
at work: a tendency to centralization on one hand and an aspiration
to liberty, autonomy and independence on the other hand. So long as
one factor checks the other there is an equilibrium and no change
is to be expected. For instance the Peruvian or the Mexican Indians
have remained for decades under the yoke of the landlords, the army
or other paramilitary groups; possible troubles
remained confined to individual villages, no collective
consciousness came into being except perhaps under some special
circumstances which lead to short-lived outbursts. On the other
end of the spectrum  is the case of a completely autonomous region;
the people then is no longer subject to any coercive control by an
alien state. The Finish Aland Islands are a possible
illustration. The move from one extreme to the other represents a
transient process; transient does not mean it should
necessarily be short; the move may take several centuries (as in the
case of Ireland) but a new equilibrium will be reached only after
this process is completed. Now, it  well known that 
non-equilibrium problems are more difficult to model than static
situations; see in this respect Herbert Simon's celebrated bowl
metaphor (Simon 1959). 
Because the
``characteristic times'' of integration phenomena are very long
(with a magnitude of several centuries) it would be desirable to
perform time averages over a period of the same magnitude. For lack
of adequate data bases this will not be possible however, and we
shall be content with time averages covering the last fifty years.
Under such conditions it will sometimes be difficult to distinguish
between quasi-equilibrium situations and transient fluctuations.

\qI{Spatial factors}

\qA{The model}

No better description can be given of what we shall for short call
Coupland's diffusion effect than the following excerpt (Coupland
1954): ``Few Englishmen want to anglicise Wales, but the pressure
exerted by the mighty neighbourhood to absorb the Welsh into the
English way of life is no less powerful because it is unconscious.
The age-long invasion never ceases; more and more English tourists
are haunting the mountains and the coasts of Wales. Aided by the
motor-car they penetrate to those districts where Welsh life has
hitherto been least affected by English contact.''
\qpar

Fig.2 schematically represents the regions in six different
countries where there have been active autonomist movements between
1965 and 1995.
\begin{figure}[htb]
    \centerline{\psfig{width=13cm,figure=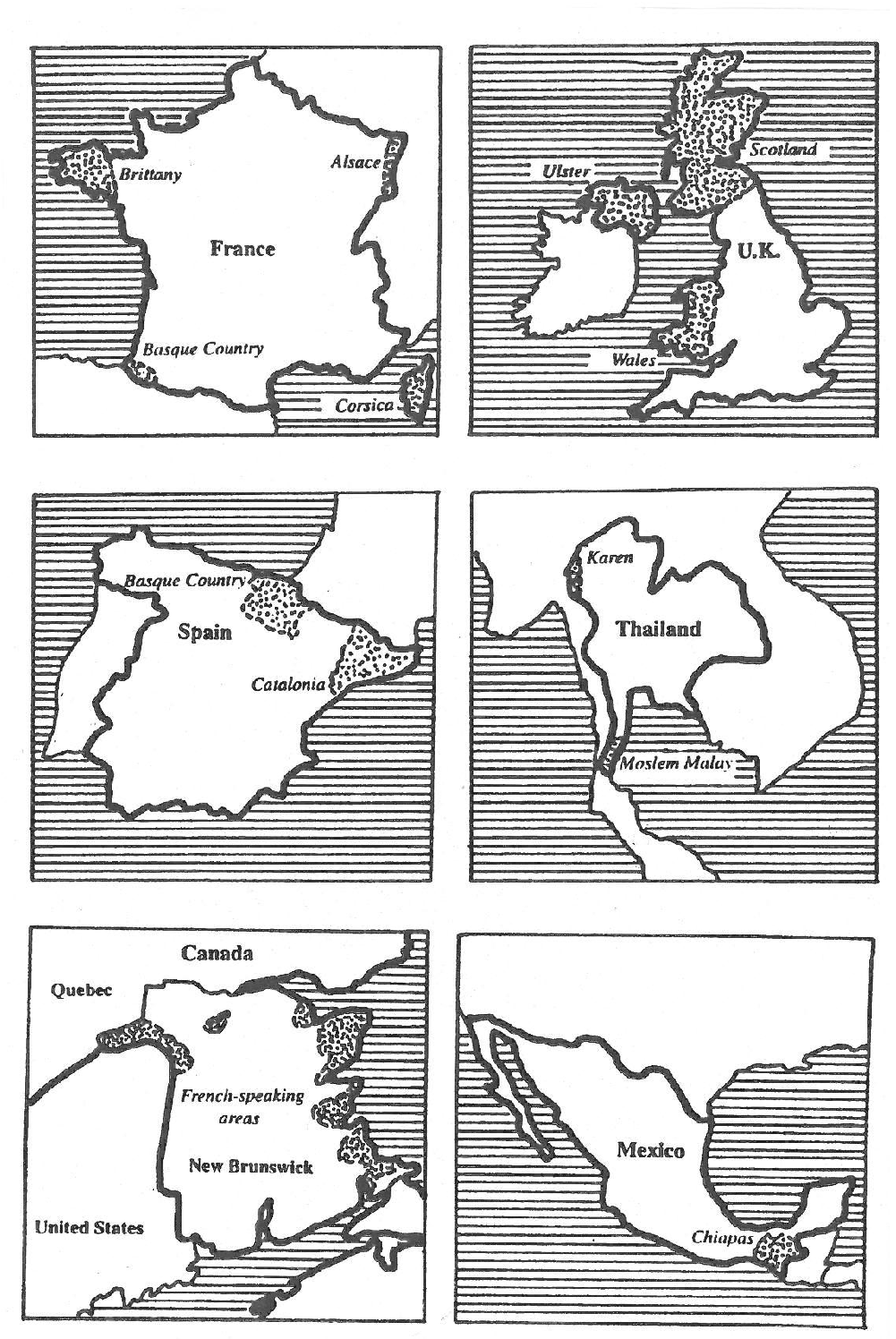}}
\qleg{Figure 2\qhv Areas with autonomist movements in 6 countries.}
{As a rule these areas
are remote from the core of the country and have minimal spatial contact
with it.Very often they are part of a linguistic entity straddling the
national borderline. For the Canadian Province of New Brunswick, the map
shows the zones where the proportion of French-speaking people is higher
than 75\%.}
{{\it Source:} For New Brunswick, Verneix (1979).}
\end{figure}
Their localization comforts the indications provided by the
Switzerland/United States comparison in section 1. Indeed all the
regions are located at the periphery of the countries in a way
which minimize the length of the borderline with the rest of the
country: two regions are islands (Corsica and Ulster), three are
peninsula (Brittany, Scotland, Wales), five are in contact with
communities which share the same minority language (Alsace, 
Catalonia, Basque Country, Southern Malay province of Thailand,
French-speaking areas in Canada's Maritime Provinces, Chiapas
state in Mexico). Together with the other observations  made in
section 1 this leads us to posit the following hypothesis.  \qpar
\qdec{{\bf The spatial determinants} of separatist tendencies in a
minority which occupies an area $ m $ in a country $ C $ can be
summarized by the following spatial integration factor: 
$$ D_s = p/lw \qn{1} $$
where:\qL
$ \bullet $ $ p $ denotes the population of the minority (in
proportion to total population). \qL
$ \bullet $ $ l $ denotes the length of the borderline between $ m
$ and $ C $ \qL
$ \bullet $ $ w $ is a phenomenological parameter representing the
facility of exchanges across the regional borderline.}
\qpar

Some additional
explanations are required regarding the evaluations of $ l $ and $
w $. 
\qpar
{\bf Evaluation of $ l $}\quad In the case of Wales, Scotland and
Brittany, $ l $ is simply the length of the borderline with the
rest of Great Britain or France. But how should we take into
account the border between Alsace and Germany for instance. The
local dialect in Alsace is a German dialect very similar to the
one spoken in Baden. Hence Coupland's diffusion argument
applies: Alsatians are able to receive German TV programs, many 
work in Germany, etc. The contact between Alsace and Germany
therefore counters the French influence; consequently that part of
the border should be counted negatively. The same reasoning holds
for the French Basque Country, the south of Thailand or for the
American Mexicans living in the vicinity of the Mexican border. The
rule for the evaluation of $ l $ may be stated in the following
form: 
\qpar
\qdec{The length of the borderline with a region
sharing the language of the minority should be subtracted from the
length of the borderline with the rest of the country; the
corresponding factors $ w_1 $ and $ w_2 $ describe the relative
strength of both diffusion processes.} 
\qpar

A further comment is in order regarding islands. Clearly it would
be inadequate to take $ l $ as being equal to zero; in the case of
Corsica for instance that would mean that there are no privileged
connections between Corsican ports and French ports such as
Marseilles or Nice which is clearly not true. We shall in such
cases take for $ l $ a conventional length which represents the
density of sea- and air- routes between the island and the
mainland.  More technical details are provided in the second
paper.
\qpar

{\bf Evaluation of $ w $}\quad In our model $ w $ is a
phenomenological parameter which can
be roughly estimated.
Broadly speaking it represents the effectiveness of transportation
and communication means%
\qfoot{Here is an example: in 1880 it took
11 hours to go from Paris to Strasbourg (500 kilometers) by train;
in 1980 it took only 4 hours; in 2010, thanks to a new high speed train
it took only 2 hours and a half.}%
. 
In other words $ w $ reflects the degree of mobilization in a
given society. For the sake of simplicity we shall restrict
ourselves to the following three-level scale:
\qee{1} $ w=1 $ for ``traditional'' economies; examples: borderline
between Biafra and Nigeria or between West and East Timor.
\qee{2} $ w=2 $ for rural economies; example: borderline between
Thailand and its southern Malay (Muslim) province.
\qee{3} $ w=3 $ for industrialized countries; example: borderline
between Quebec and the rest of Canada.

\qA{Application to the comparison: Switzerland/United States.}

By way of illustration let us apply our previous definitions to the
comparison of the process of integration in the United States and
in Switzerland that we already discussed in section 1. The
parameter $ w $ is the same for both countries  and  it can
therefore be dropped for the purpose of comparison. In the case of
Switzerland $ l $ and $ p $ have remained almost unchanged during
the last centuries: $ l \simeq 200\ \hbox{km},\quad p = 20\% $. On
the contrary in the United States, the situation of the people of
German ancestry has changed drastically in the course of time;
therefore a specific date has to be selected for the evaluation of
the spatial integration factor $ D_s $; let us take $ 1900 $. The
exact geographical repartition of the American Germans is not known
 (most certainly it would show a fractal pattern); this does
not matter however for an educated guess is sufficient for our 
purpose. The American German community numbered about 14 million
(this figure has been extrapolated from the proportion of people of
German ancestry given in subsequent censuses); this leads to $
p=15\% $ . Let us consider that, being mainly an urban population,
it was distributed in urban groups of about 50,000 people; this
gives about 320 groups. Let us consider that the density in such
cities was of the order of 2,500 inhabitants
per square kilometer. The  length of the borderline of one group
with the rest of the country thus turns out to be equal to $ 15.8\
\hbox{km} $ and for the total borderline we get: 
$ l = 320\times 15.8 = 5,056 \
\hbox{km} $. As a consequence  we get the following integration
factor:  
$$ \matrix{
 & \hbox{USA (1900)} & \hbox{Switzerland} \cr
 & \hbox{\small German-speaking Americans} &\hbox{\small French-speaking Swiss}
\cr 
\hbox{ Spatial integration factor} (D_s)  \hfill & 25,000 & 600 \cr
} $$

Thus, the integration pressure thus was 42 times larger in
America than in Switzerland, an observation which is of course
consistent with subsequent evolution.
To conclude our
demonstration it remains to document our assertion about the poor
level of  integration  in Switzerland. Two important elements are
provided in Fig.3. 

\begin{figure}[tb]
    \centerline{\psfig{width=16cm,figure=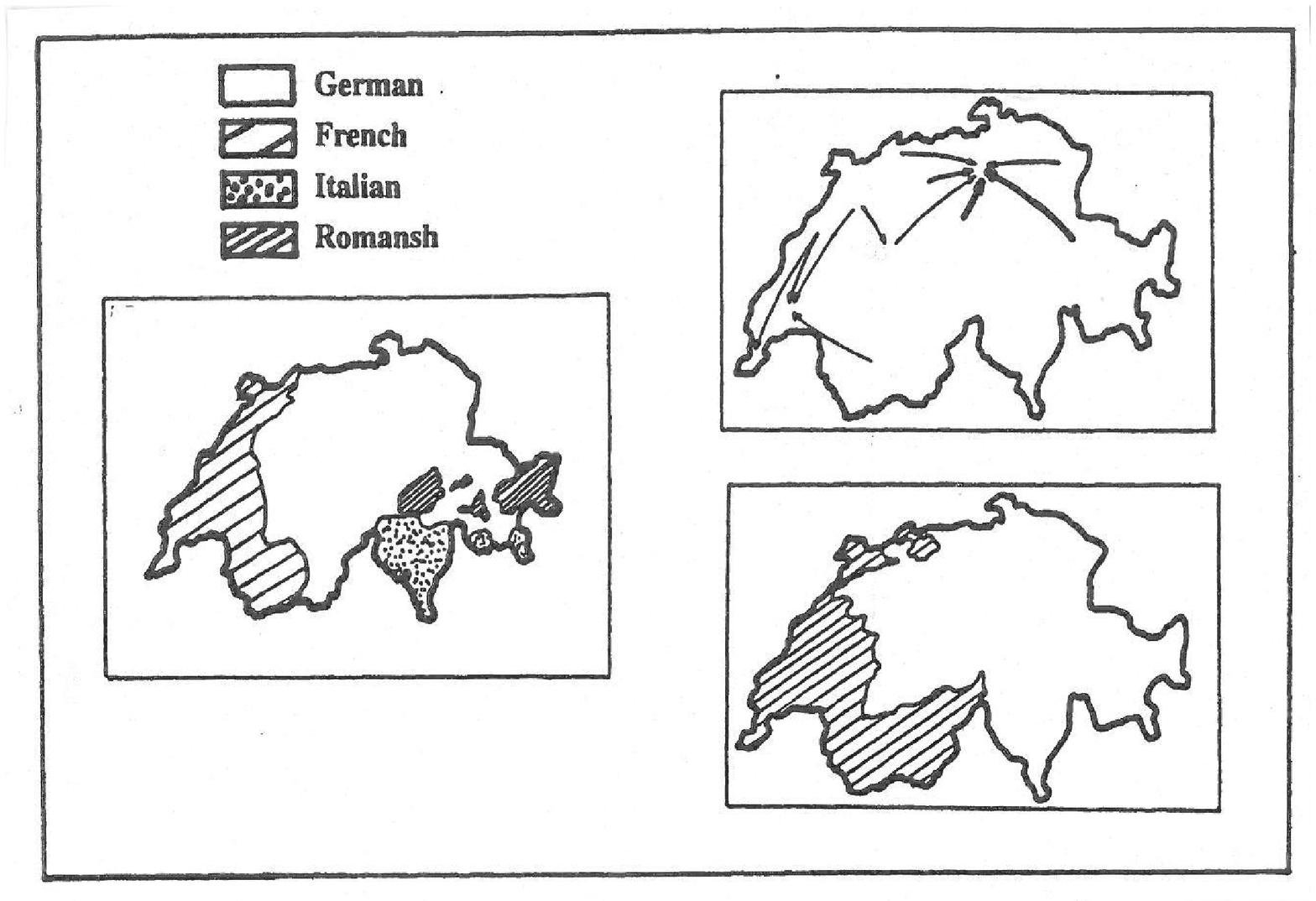}}
\qleg{Figure 3\qhv  Linguistic compartmentalization versus
migration and political choices.} 
{The map on the left-hand side shows the
linguistic zones in Switzerland (1980). The map in the upper right corner
shows the net migration between different so-called employment regions
1975--1980. The conclusion which can be drawn
is that there are no substantial
migration flows across linguistic frontiers. The map in the lower right
corner shows the results by canton of the referendum of 6 December 1992
on the project of European Union membership. A majority of voters in the
hatched zone voted yes; the results  closely followed linguistic
subdivisions. A more precise analysis would require electoral results at
the communal rather than at the cantonal level, especially for Valais
(which is French-speaking in the proportion of 60 percent).}
{{\it Sources:}
Bassand et al. (1985), Jenkins (1986), Racine (1994).}
\end{figure}
%
The first one shows that there is a very
small residential mobility between the French- and the
German-speaking  parts; the second shows that there is a
major fracture between the two regions regarding a problem of 
cardinal importance for the future of the country. One could add
that the French- and Italian-speaking minorities are under
represented at the level of the high federal administration:  
$$ \matrix{
 & \hbox{Percentage of the minority}  & \hbox{Percentage in the} \cr
 & \hbox{in total population} & \hbox{high federal administration}
\cr
 \hbox{French-speaking minority} \hfill & 20\% & 12.5\% \cr
\hbox{Italian-speaking minority} \hfill & 6\% & 2.7 \% \cr
} $$

{\it Sources: Gazette de Lausanne (June 6, 1955); Le Monde (May 29,
1978)} 
\qpar

 Moreover, in the
German-speaking part of Switzerland the usage of Swiss-German, once
restricted to domestic and casual communication, is becoming more
and more widespread even for instance in the classroom or in
advertisements. As a result, a French- or Italian-speaking Swiss
would have to learn three languages in addition to his own mother
tongue, namely: English (imperative in the present world), German
and Swiss-German. This is a very demanding task  and it
is doubtful that it could be carried out
successfully by a large proportion of the population. 

\qA{Discussion of apparent counter-examples}

Before examining to what extent the previous model is supported by
statistical evidence, it is perhaps worthwhile to discuss what
could seem to be obvious counter-examples. 

\qun{The American Civil War}
The most spectacular
case is the American Civil War. In French it is referred to as the
``Guerre de S\'ecession'' [Secessionist war]
but was it really  a secessionist struggle of the
kind considered in this paper? One element, the attachment to a land
that has been occupied for generations, was certainly present
in the South. Yet, neither language nor religion was much
of an issue.   In other words we have here the worrying situation
of an allegedly  separatist struggle of first magnitude (600,000
deaths, Richardson 1960) that seems to fall outside the scope of our
model. Our contention is that it was {\it not} a separatist
struggle of the kind analyzed in this paper;
 The
matter would certainly deserve a detailed discussion in its own;
let us here restrict ourselves to the following, schematic
arguments.  
\qee{1} In an ethnic conflict
defeat does not mark the end of the struggle. A comparison with
other wars that occurred at about the same time will illustrate
the point. In 1830 the first Polish rebellion was suppressed
without mercy (16,000 deaths); in 1848-1849 the Hungarian uprising
met the same fate  (1,500 deaths). Yet, both the Polish and the
Hungarian engaged in other attempts until eventually they won their
independence.  
\qee{2} How then, if it was not a separatist struggle,
should the American Civil War be categorized? 
It was the last fight
of a minority (5.5 million against 22 million) for the defense of
its way of life and its political autonomy; in other words it was a
war for political leadership, not  for ethnic survival. More or less
of the same type are the wars of Vend\'ee in Revolutionary France
or the war between Bavaria (allied to Austria and other German
states) against Prussia (1866); the latter can be considered
as the last feat%
\qfoot{True, there have been 
other short-lived separatist attempts in Bavaria (1918,1923), but
none that was of much consequence.}
 of a sovereign nation before
accepting its absorption in a larger entity. Incidentally, it can
be noted that the magnitudes of the respective populations were
about the same  than for the American Civil War: Prussia: 20
million, Bavaria: 5 million. 

\qun{The Amish}
The next example is more embarrassing. The Amish are a small
American community which has maintained its faith {\it and its
language} (a German dialect) since the time they arrived in America
in the 19th century. Let us first estimate the magnitude of the
integration factor for the Amish as compared to that for the
entire American German community: in 1900, if we assume that the
Amish settled in urban communities of the same size as the
average German Americans, we get the same  integration factor 
(indeed this factor turns out to be independent of the size of the
minority under consideration). In fact two corrections should be
introduced: firstly,  
given their rural way
of life it is  more reasonable  to assume that the Amish settled in
smaller communities than other German Americans; this would result
in a larger integration factor than for German
Americans; secondly, since the Amish conserved their nineteenth
century way of life (no cars,  electricity,  radio, or television)
the value of their mobilization factor $ w $ did not increase in
the same way as that of other American Germans; this would  reduce
the value of the integration factor, at least for the twentieth
century. By and large it is probably not unreasonable to admit
that these increase and decrease approximately cancel each other.
Consequently, if  the integration factor is to be taken at its face
value the Amish should have lost their mother tongue in the same
conditions as other American Germans.  Yet, that did not happen.
Why? 
\qpar
Two elements of answer can be given.  
\qee{1} The Amish constitute a special case in the
sense that they managed to establish a strong and durable connection
between their religion, their way of life  and their language. Not
surprisingly, they had
no difficulty in maintaining their faith, but because of the close
connection between their faith and their language the latter was
maintained too%
\qfoot{This ``explanation'' merely displaces the
problem: why indeed did the Amish succeed in maintaining such a
close link between faith and language? The answer is not obvious.
The Mormons for instance once tried to introduce a special writing;
but for all their organizational skills the tentative did not meet
with much success and had eventually to be abandoned.}%
. 
\qee{2} There is in fact a substantial emigration rate out of the
Amish community; but because of fairly high birth rates, this does
not imperil the community's future. 

\qA{The integration process at parochial level}

Why did social mobilization often stimulate nationalist feelings
instead of curtailing them? Such an observation has been made
repeatedly  (communal revivals in Western
Europe are obvious examples) and it seemed to challenge K.
Deutsch's mobilization thesis. This was all the more puzzling
because  the mobilization thesis
had a firm, clear and convincing basis at the 
microsociological level; how could increased
interaction  may have the effect of seemingly reducing integration?
One explanation to this paradox consists in distinguishing between
several processes of integration taking place at different spatial
levels. About one or two centuries ago a region of
the size of Brittany did not constitute a homogeneous ethnic
entity. In the case of early nineteenth century France, E.
Weber (1976, p.45) notes that ``the least of the village considered
itself a specific homeland in its language, legends, customs,
ways''. He shows persuasively that acute parochialism was a general
``disease'' in most French provinces of that time.
For instance real hostility existed between Lower and Upper
Brittany, a division based on the  Celtic speech of the former and
the French dialect of the latter. Very often, either in Brittany or
in other French provinces, local ``wars'' broke out between
different parishes that gave rise to real (albeit small) battles,
most of which did not leave any record in historical archives
(see however the ``Statistiques du d\'epartement du Lot'' 1831).
Similarly in its systematic analysis of French rebellions, C.
Tilly (1986) came to the conclusion that ``broadly speaking, the
repertoire of the mid-seventeenth to mid-nineteenth centuries has 
a parochial scope. The repertoire that crystallized in
the nineteenth century and prevails today is, in general, more
national in scope.''
 In
Ireland too the divergent interests of various clans (the O'Rourke,
the O'Donnel, etc) undermined any permanent coalition%
\qfoot{Even
today, the nationalist fight of the Kurds is hampered by a
chronical lack of solidarity between the different ethnic
components of the Kurdish people (Premdas et al, 1990.}%
.  
Thus, a precondition for an effective nationalist struggle was the
completion of a reasonable degree of regional integration. 
This conclusion has been summarized in the following way by C.
Tilly in a recent work (1993): ``During the 16th and 17th
centuries the effective units of collective action in Ireland
consisted largely of patron-client chains led by warlords.
Only during the 19th century, as class-coalition and
national revolutions were generalizing elsewhere in Europe, do we
see a popularization of the Irish cause at a national level.''
In
short one of the keys to the solution of the problem under
consideration is the fact that (at least%
\qfoot{At the level of capital cities a third process
of international integration could be mentioned}
)
two integration processes were under way simultaneously: 
\qbu a national integration which concerned the urban
components of the society
\qbu a regional integration which concerned rural areas,
especially those with inadequate transportation networks.

\qA{Respective roles of religion and language loyalties}

The spatial part of our model is specifically designed to account
for the phenomenon of language maintenance. This may seem to be a
serious restriction. In fact, as will be shown in this paragraph,
this limitation is not so unmitigated as it could
seem at first sight. We shall argue that,
before the twentieth century, religions very much played
the role languages have today in establishing national
loyalties. 
Today, especially in industrialized countries,  the mechanisms
of religion and language maintenance are very different.
Economic competition in the employment market produces a very
effective incentive to speak the language of the dominant
majority. Nothing of the sort exists for religions. As a result,
as far as religions are concerned, the assimilation rates are
expected to be much lower than for languages, intermarriage 
being in this case the main ``assimilation'' mechanism.
Two or three centuries ago the
situation was completely different for the very definition of
citizenship was based on religion. This was not only
true in Catholic or Muslim states, but also in those protestant
states such as Denmark, Great Britain or Sweden which are usually
considered as having been more tolerant in religious matters.  Let
us illustrate our proposition by three examples:  
\qbu In
1709 when the Whigs gained a clear majority in Parliament they
passed an act that provided a simple procedure for the
naturalization of foreigners. Aliens had only to swear allegiance
to the Crown, prove they had received a Protestant sacrament in the
preceeding three months and declare in open court against the
doctrine of transubstantiation. A similar law was enacted in 1740
for the purpose  of permitting foreigners in America to acquire
subjectship (Thernstrom 1980).
Similarly, by the Test Act (1673), every civil servant had to
declare against the doctrine of transubstantiation (Mourre 1978,
vol.1, p.246). It is in 1679 that the Catholics were explicitly
excluded from the Parliament.
Reference to the Christian faith was maintained until the late
nineteenth century in the oath taken by  members of the Parliament:
it is only in 1866 that the expression ``upon the true faith of a
Christian'' has been deleted (Doubnov 1933).
 \qbu In Denmark, the civil rights of the catholic minority
had been severely curtailed by successive laws enacted in 1613,
1624, 1643; they were confirmed and strengthened by Christian V
after his accession to the throne in 1660. In Sweden 
the discrimination against catholics, introduced by Charles X, was
extended by Charles XI (1660-1697): conversion to the catholic
religion was forbidden under pain of banishment and deprivation of
all possessions (Daniel-Rops 1958)
\qbu Our last example concerns Germany; in the numerous little
kingdoms that composed Germany before 1870 the prevailing rule was 
``cujus regio, ejus religio'' (the religion of the sovereign
determines the religion of the citizens). 
 \qpar

One should also recall that education, which has become so
important an issue in modern societies, was closely connected with
religion until the end of the nineteenth century. One of the major
purposes of elementary education was to be able to read the Bible.
This was particularly true in Protestant or Presbyterian regions.
In the case of Wales for instance, R. Coupland (1954) relates the
remarkable development of the so-called ``Sunday schools'': ``On
Sunday the whole country was turned into a religious school; hour
after hour these Welsh folk sat listening to the reading of the
Bible and the geography of Palestine became more familiar to them
than that of Wales itself''.\qL
A last confirmation of the shift from religions to languages
issues can be gained from the example of the Bernese Jura in
Switzerland. Between 1830 and 1950 there have been five major
outbreaks: 1834, 1867, 1873, 1917, 1947. The first three had
religious mobiles, while the last two had linguistic causes;
needless to say, the movement that began after 1965 and lead
eventually to the creation of a new canton was largely based on
claims for linguistic rights. 
\qpar

Communal revivals in western industrialized  societies may often
be traced back to the survivance of religious traditionalism. In
such cases as Ireland or even Wales the role played by religion is
well known. Perhaps less known but quite as revealing are the
origins of regional revivals in France. The proportion of the
people attending Mass is notably higher than average in all the
regions (the only exception being Corsica) where there has been a
communal revival (Isambert and Terrenoire 1980,p.30; the figures
refer to the period 1955-1962): 
$$ \matrix{
\hbox{Alsace} \hfill & 84\% \cr
\hbox{Basque Country} \hfill & 79\% \cr
\hbox{Brittany} \hfill & 67\% \cr
\hbox{\bf National average} \hfill & \hbox{\bf 35\%} \cr
\hbox{Corsica} \hfill & 25\% \cr
} $$

In conclusion one may say that, seen in historical perspective, the
lack of integration pressure regarding religious matters is  a very
recent innovation in industrialized societies. In most developing
countries religion  still is a key factor in
determining social and professional loyalties; besides, there is
probably a much greater resemblance between the mechanisms
governing religion or language maintenance, 
 than may be suggested by the example of modern
western societies.

\qI{Semi-quantitative tests}

\qA{Official languages}

It may be of interest to examine to what extent each
sovereign country has its own distinctive official language. At
first glance it may seem that many nations share one of the world's
dominant languages. To begin with, it should be noted that, in
contrast to an overwhelming majority of nations, neither the United
States (at least at the federal level) nor the United Kingdom have
an official language. Furthermore, the official language of many
former colonies still remains the language of the former colonizing
power. In addition it is well known that many nations have Arabic
as their official language. In order to assess the extent of these
effects one has to look at the data in some detail. The results are
summarized in Table 1 based on Banks (1995) and the ``Europa World
Yearbook'' (1995). 
\qpar
$$ \matrix{
\hbox{Number of sovereign nations} \hfill & 193 & \cr
\hbox{Countries having at least one specific national language}
\hfil & 92\ & 48\%  \cr
\hbox{Former colonies (independent after 1945)} \hfill  & 46 & \cr
\hbox{Countries having at least one specific national language}
\hfill & &   92/(193-46)= 63\% \cr
\hbox{\hfill (former colonies excluded)} & &   \cr
 & & \cr
\hbox{Spanish as official language} \hfill & 15 & \cr
\hbox{Arabic as official language} \hfill & 17 & \cr
} $$
\centerline{{\bf Table 1: Official languages}}
\qpar

It should be noted that there is a marked, albeit slow, tendency
especially for former colonies to adopt proper mother tongues as
official languages. Examples are: India (English is no longer an
official language although it still has the status of an
associate language for many official purposes), Luxembourg
(Letzeburgish,1982), Kenya (Kiswahili,1974), 
Seychelles (Creole,1981).

\qA{Island states}
The lack of integration can manifest itself in two ways: in the
existence of regional conflicts between a minority and a dominant
majority, or in the existence of independent  states, the latter
being the obvious outcome of successful separatist struggles. In
our model islands play a special role since, because of their
minimal contacts with neighboring nations,they are 
characterized by a particularly small integration factor. 
\qpar
 
{\bf A case study: the Aland islands} \quad
Before we undertake
 a systematic analysis let us first consider a special
case, namely the Swedish minority in mainland Finland on one  hand
and in the Aland islands (which belong to Finland) on the other
hand. In Finland a slow process of assimilation has been under way
for decades  as illustrated by the following data giving the vote
for the Swedish People Party (Svenska Folkpartiet); the fall in the
vote $ v $  can be modeled
by: $ v = e^{-s/0.75} $ with $ s $ being expressed in centuries:  
$$ \matrix{
         & 1907 & 1917 & 1927 & 1936 & 1948 & 1958 & 1966 & 1979 &
1987 \cr
\hbox{SPP } (\%) \hfill & 12.7 & 10.9 & 12.2 & 11.2 & 7.7  & 6.7  & 6.0  &
4.6 &   5.3 \cr
} $$

$$ \matrix{
     & \quad 1880 \quad & \quad 1950 \quad \cr
\hbox{Swedish-speaking people } (\%) \hfill & 14.3 & 8.6 \cr
} $$

{\bf Table 2: Votes for the Swedish People Party and
proportion of Swedish-speaking people.} {\it Sources: Mackie and
Rose (1982); Jansson (1961)}
\qpar

The situation is completely different in the Aland islands. Its
population of about 21,000  is  Swedish-speaking at 95 percent. 
Ceded to Russia in 1809, the Islands were returned to Finland by
the League of Nations in 1921 despite calls by the islanders for
reunification with Sweden. The islands have been demilitarized
since 1856 and are neutralized since 1921. In exchange of the
engagement of the
 Finnish government to respect and preserve the
Swedish language of the islanders Sweden agreed to
withdraw its claim to sovereignty over the islands. Since that
time, the islanders maintain a loyal attitude toward Finland but,
in contrast with the mainland Swedish-speaking minority, there has
been no further progress of their integration in the Finnish
community. 
\qpar

Let us now examine the relationship between insularity and
aspiration to sovereignty in a more general way. 
 In the case of an island, the
separatism factor is particularly high; accordingly one expects
a large number of island-states. This effect should be
particularly marked for small states since there are only few
large islands (of the size of Britain or Japan for instance). 
In order to carry out our test we
selected a threshold area of 14,000 square kilometer. While being
of course somewhat arbitrary, this limit roughly corresponds to the
category of what may be called ``small countries''; as a matter of
comparison, Belgium has an area of 11,778 square kilometer and
Lebanon of 10,452 square kilometer. 
 At first sight, especially from an European perspective,
our conjecture may be found to be wrong; there are indeed quite a
large number of small sovereign countries which are {\it not} islands:
Andorra, Belgium, Liechtenstein, Luxembourg, Gibraltar, Monaco,
Vatican. For the whole world, however, our conjecture is
confirmed: there are 47 states of less than 14 000 square
kilometer, of which 35 are islands, a proportion of 0.74
(Statesman's Yearbook 1995).

\qI{Historical factors}

\qA{The model}

As has been persuasively argued by Connor (1972, 1994) or Jenkins
(1986) history plays a fundamental role in shaping the collective
psychology of peoples. For instance, if a nation has had the
status of a sovereign state for a long time it will not easily
accept to be incorporated in an alien state; Lithuania and Poland
are illustrative examples. Similarly, if the conquest of a region
has required a protracted and costly war, there is a high
likelihood of subsequent rebellions: Algeria, Chechniya and Ireland
are illustrative examples. Even once a nation has been subdued, the
aspiration to sovereignty will remain strong for centuries. We
shall generalize this kind of observation by assuming that those
attitudes or actions that have been resorted to several times in
the past will tend to repeat themselves in the future. This, in
fact, has become a fairly standard assumption in a number of
different contexts as we shall see now.
\qee{1} Drawing upon a formidable array of empirical evidence,
C. Tilly (1986, p.390) introduced the conception of a
people's repertoire of action: ``Any population has a limited
repertoire of collective action. People tend to act
within known limits, to innovate at the margins of existing forms.
People know the general rules of performance and vary
the performance to meet the purpose at hand.''
\qee{2} In the field of operational research our assumption is
embodied in the observation that, at least in the start-up phase
of the learning curve, the completion of a fairly complex task
requires less and less time and effort as it is performed again
and again; typically, it takes only half as much time to perform a
task for the 10th time than it took for the first time (Baloff
1971, Hamblin et al 1973). 
\qee{3} In the field of history, it has been showed (Roehner
1993a,b) that the above principle plays a role in a wide range of
historical actions. For the sake of brevity it has been referred
to as the {\it paronymy assumption} (collusion and collision, or
gradation and graduation are paronymic words). In mathematical
terms it can be stated in the following form: 
\qdec{\it The probability
of a given action is in proportion to the number of its former
occurrences.}
\qpar

In the present study the paronymy assumption will play a dual
role: (i) It provides a guide to the way empirical evidence should
be recorded and aggregated; more specifically, we shall
distinguish two periods: 1850-1945 on one hand, 1945-1995 on the
other; and we shall examine if there are any similarities in the
means used by separatist movements in those periods. (ii) By
comparing the ways of separatist movements before and after 1945
we shall also be able to test the paronymy assumption. This will
be done here in a rather qualitative way by examining three
examples; in the second part of this study such tests are
carried out in a more systematic way. 
 
\qA{The Bernese Jura in Switzerland}
Our analysis relies on two main sources: Jenkins (1986) and
Rennwald (1984). Before and after 1945, separatism from the
Bernese canton manifested itself in very similar ways: petitions,
opposition displayed in general elections or in referendums,
occasional outbursts of violence which brought about the deployment
of Bernese troops. The only
major innovation in the post-1945 period was the occurrence
of bombing (at least the limited information that we have on the
pre-1945 period does not mention such episodes). That innovation
may largely  be due to technical reasons which made bombs
easier to store and to handle; this may explain why bombing 
has been
used by almost all  separatist movements in Western Europe. The high
degree of historical continuity may also be emphasized by the fact
that the border of the new Jura canton follows almost exactly the
limit between the catholic and the protestant subregions as
recorded in the Treaty of Aarberg of 1711 (Jenkins 1986). \qL 
A
similar analysis could be conducted for a number of other regional
movements in industrialized countries: Alsace, Belgium, Brittany,
Scotland, Wales. The French Basque Country and Corsica are to some
extent exceptions in the sense that there is no record of any
significant and organized separatist movement between 
1850 and 1945%
\qfoot{These movements can be considered as
illustrations of the ``demonstration effect'' (Connor 1994); they
may have been been triggered by the separatist rebellion in the
Spanish Basque Provinces }
Yet, in both regions the forms taken by separatist
movements is consistent with their respective ``traditions'': in
the Basque Country it relied heavily on  religious feelings and
institutions, in Corsica it was channelled by the long-standing
vendetta tradition.  

\qA{Aceh province (northern Sumatra, Indonesia)}

Our analysis relies on two main sources: Cribb (1992) and Zainu'ddin
(1968). The record here is very different from that for the Bernese
Jura: warfare and guerilla tactics are the main forms of
separatist struggles in the Indonesian province of Aceh.
According to official estimates the thirty-year war 
(1873--1908) waged by the Acehnese against the Dutch cost 100,000
deaths on the Acehnese side and 12,000 on the Dutch side. It is
labelled as having been a ``ferocious'' war although no detailed
historical record seems to  be available. A similar pattern
occurred once again under Indonesian (Javanese) occupation. In
September 1953, there was an uprising lead by the former
military governor Daud Bereuh. Until 1959 the region remained
almost completely independent from Djakarta. A new separatist
uprising broke out in 1990; within a single year the toll was
estimated to about 2300 deaths. In July 1993, ``Amnesty
International'' denounced the violation of human rights by the
Indonesian army. A similar analysis could be conducted for a 
number of other
regional movements in developing countries, for instance: for
instance Jolo and Mindanao in the Philippines, the Sikh separatist
movement in India, the struggles for Indian rights in Mexico
(Chiapas), in Peru or in Bolivia. In many of these instances it is
difficult to find detailed historical records for the
 period before World War II; from the few records that are indeed
available one gets the impression of a strong historical
continuity.
 Yet, it seems that there also a few
exceptions: for instance the Maori struggle in New Zealand has not
resumed the warfare form it assumed in three nineteenth century
Maori wars. These
exceptions will be examined more closely in the follow-up paper.  

\qA{Basque Provinces versus Catalonia (Spain)}
This case has captured the imagination of a number of
researchers because the Basque movement is marked by
terrorism while political negotiations prevailed in Catalonia. To
explain this contrast several mechanisms have been proposed which
are reviewed in Laitin (1995)%
\qfoot{Let us emphasize that it would be pointless to
propose a specific ``explanation'' for this  single case; even if
it turn out to work in that case, which is assured so to say by
construction,  it will be nothing but a
tautology if it
is unable to explain other cases as well. D. Laitin is well aware
of this pitfall since he tests his own model on a second case,
namely the contrast between Georgia and Ukraine.}%
.
Let us see if there is a definite answer to this paradox
 in the general framework of the paronymy assumption. Our
contention is that throughout their recent history Basques were
more prone to resort to political violence and military solutions
than were the Catalans. 
Let us
limit ourselves to discussing how both peoples reacted to and took
part in the many internal conflicts that affected Spain during the
19th and 20th centuries. There have been five main wars: the
four so-called Carlist wars (1833-1839, June 1848--April 1849,
1860, 1872--1874) and the Spanish Civil War (1937-1939). Let us
examine them in turn. 
\qee{1} Barres du Molard (1842) who was Charles V's chief of staff
gave a detailed account of the battles that have been fought
in the First Carlist War (1833--1839). Out of 63 battles, 59 took
place in the Basque Provinces (Navarra included) while 4 took
place in Castille. In spite of  a substantial amount of social
agitation in Catalonia no battle was fought there. 
\qee{2} The Carlist attempts which took place in 1848--1849 and
in 1860 were not very serious ones. In 1848, Alzda took command
of the Carlists in the Basque Provinces; he was captured and shot
(Clarke 1906, p.214). At the same time, Cabrera entered
Catalonia through French Cerdagne (June 1848); he spent almost
one year there  without any significant fighting taking place; in
April 1849 he abandoned hope and crossed the frontier again. In
April 1860, Don Carlos Luis landed with 3,500 troops near Tortosa
(mouth of the river Ebro). But the expected rising did not occur
and he was captured without putting up any resistance.
\qee{3} The protracted war of 1872--1874 was marked by the
following major battles: Orioquieta (Navarra, May 1872), Estella
(Basque Provinces, August 1873), Tolosa (Basque Provinces,
November 1873), Vich and Olot (Catalonia, January and March 1874),
Teruel and Cuenca (Castille, July 1874), siege of Bilbao and Irun
(Basque Provinces, 1874). Thus, out of 9 battles only 2 took place
in Catalonia. 
\qee{4} During the Spanish Civil War, the nationalists first
subdued the Basque Provinces. From September 1936 to August 1937
nationalist forces proceeded from East (San Sebastian) to West
(Santander). The campaign that took place in Catalonia (December
1938--February 1939) was one of the last phases of the war and was
much shorter than the struggle in the Basque Provinces
\qfoot{The latter is said to have cost 50,000 deaths (Davant
1975); we were  unable to find a corresponding figure for the
war in Catalonia.}
.
\qpar

Broadly speaking, on the basis of the previous historical record,
one may say that the repertoire of the Catalan people, although it
included different varieties of general strikes,  urban
uprisings or self-promulgated autonomy proclamations was rather
poorly endowed in terms of stubborn fighting or military
upheavals. This may account for the difference between Basque and
Catalan autonomist movements after 1960. Such ways manifested
themselves once again in the autonomy movements of the 1980s.

\qI{Conclusion}

We have been interested in ethnic revivals based on homeland
loyalties and in which linguistic factors usually play a prominent
role. We presented a model aimed at describing the long-range
trend of such phenomena. In a long-term perspective, spatial and
geographical factors turn out to be of cardinal importance. In
this paper our objectives were: (i) To describe the mechanisms we
believe to be at work (ii) To discuss the validity and the
implications of the model on some critical examples (iii) To
emphasize that in the emergence of national loyalties there is a
strong parallelism between the role  once held by religions and
the one languages play nowadays. 
In the second part of this paper we examined how and to what
extent the historical legacy determines and shapes 
nationalist struggles. 
The present model has a number of straightforward
implications. As an example let us come back to the
integration process in the United States whose
strength and effectiveness we emphasized in section 1. There are
only two countries contiguous to the United States. Canada being an
English-speaking country (with the obvious exception of Quebec)
there can be no major minority problem on the northern frontier. On
the basis of history and of language quite a different situation
would be expected to prevail on the southern border with Mexico, a
surmise which is indeed confirmed by the observation (Lieberson et
al 1975, p.56) that in New Mexico only a small proportion of the
Spanish-speaking people has shifted from Spanish to English since
1846. 
\qpar

In the follow-up paper we shall subject our
model to more systematic tests using a data set based on 40 cases
(over the period 1945--1995). Yet, as pointed out by K. Popper,
no finite set of tests can ever validate a theory completely;
ultimately, it is the absence of contradictory evidence which
becomes proof of the theory. It is this very argument which gives
its significance to the case studies reported in the present paper. 

\vfill \eject




\centerline{{\Large References}}
\vskip 1cm

\def\qparr{ \vskip 1.0mm plus 0.2mm minus 0.2mm \hangindent=10mm
\hangafter=1}

\qparr
ALLARDT (E.) 1979: {Implications of the ethnic revival in
modern industrial society. A comparative study of the linguistic
minorities in Western Europe.} Societas Scientiarum Fennica.
Helsinki.

\qparr
BALOFF (N.) 1971: Extension of the learning curve. Some empirical
results. Operational Research Quarterly 22,4,329-340.

\qparr
BANKS (A.S.) ed 1995: Political handbook of the world 1994-1995.
C.S.A. Publications. Binghamton (N.Y.)

\qparr
BARRES DU MOLARD (A.) 1842: {M\'emoires sur la guerre de la
Navarre et des Provinces basques depuis son origine en 1833
jusqu'au trait\'e de Bergara en 1839.} Dentu. Paris. 

\qparr
BASSAND (M.) et al 1985: {Les Suisses entre la mobilit\'e et
la s\'edentarit\'e.} Presses Polytechniques Romandes. 

\qparr
BELL (W.) 1954: A probability model for the measurement of
ecological segregation. Social Forces 32 (May),357-364.

\qparr
BRUN (A.) 1946: {Parlers r\'egionaux. France dialectale et
unit\'e fran\c caise.} Didier. Paris.

\qparr
CAMILLERI (G.), LAZERGES (C.) 1992: {Atlas de la criminalit\'e
en France.} Reclus-Documentation Fran\c caise. Paris. 

\qparr
CLARKE (H.B.) 1906: {Modern Spain 1815-1898.} Cambridge
University Press. Cambridge. 

\qparr
CONNOR (W.) 1972: Nation-building or nation destroying. World
Politics 24,3,319-355.

\qparr
CONNOR (W.) 1994: {Ethnonationalism. The quest for
understanding.} Princeton University Press. Princeton.

\qparr
COUPLAND (R.) 1954: {Welsh and Scottish nationalism.} Collins.
London. 

\qparr
CRAWFORD (J.) ed 1992: {Language loyalties. A source book on
the Official English controversy.} University of Chicago Press.
Chicago.

\qparr
DANIEL-ROPS 1958: {Histoire de l'Eglise. Le grand si\`ecle des
\^ ames.} Fayard. Paris. 

\qparr
DAVANT (J.L.) 1975: {Histoire du Pays Basque.} Editions
Elkar. Bayonne. 

\qparr
DEUTSCH (K.W.) 1979: {Tides among nations.} The Free Press.

\qparr
DEUTSCH (K.W.) 1984: Space and freedom. Conditions for the
temporary separation of incompatible groups. International
Political Science Review 5,2,125-138.

\qparr
DOUBNOV (S.) 1933: {Histoire du peuple juif.} Payot. Paris. 

\qparr
EUROPA WORLD YEARBOOK 1995: Europa Publications Limited. 

\qparr
FISHMAN (J.A.) 1986: Bilingualism and separatism. Annals of the
American Academy of Political and Social Science 487,169-180.

\qparr
GOTTLIEB (G.) 1993: {Nation against the state: a new approach
to ethnic conflicts and the decline of sovereignty.} Council of
Foreign Relations Press. New York.

\qparr
GRIMMET (G.R.), STIRZAKER (D.R.) 1982: {Probability and
random processes.} Clarendon Press. Oxford.

\qparr
GURR (T.R.) 1993: Why minorities rebel: a global analysis of
communal mobilization and conflict since 1945. International
Political Science Review 14,2,161-201.

\qparr
HAMBLIN (R.L.), JACOBSEN (R.B.), MILLER (J.L.L.) 1973: {A
mathematical theory of social change.} John Wiley. New York.

\qparr
HENDERSON (J.L.), CALDWELL (M.) 1968: {The chainless mind. A
study of resistance and liberation.} Hamilton. London.

\qparr
HORN (N.), TILLY (C.) 1986: Catalogs of contention in Britain,
1758-1834. New School for Social Research. Working Paper No32.

\qparr
HOROWITZ (D.L.) 1985: {Ethnic groups in conflicts.} University
of California Press.

\qparr
JANSSON (J.-M.) 1961: Bi-lingualism in Finland. Fifth World
Congress of the International Political Science Association
(September 26-30).

\qparr
JENKINS (J.R.G.) 1986: {Jura separatism in Switzerland.}
Clarendon Press. Oxford.

\qparr
KHONG (Y.F.) 1992: {Analogies at war. Korea, Munich, Dien
Bien Phu and the Vietnam decisions of 1965.} Princeton University
Press. Princeton. 

\qparr
KURIAN (G.) 1979: {The book of international lists.} Macmillan.

\qparr
LAITIN (D.D.) 1995: National revivals and violence. Archives
Europ\'eennes de Sociologie, 26,3-43. 

\qparr
LEACH (E.R.) 1954: {Political systems of highland Burma.}
Harvard University Press. Cambridge.

\qparr
LIEBERSON (S.) 1970: {Language and ethnic relations in
Canada.} Wiley. 

\qparr
LIEBERSON (S.), HANSEN (L.K.) 1974: National development, mother
tongue diversity and the comparative study of nations. American
Sociological Review 39,523-541.

\qparr
LIEBERSON (S.), DALTO (G.), JOHNSTON (M.E.) 1975: The course of
mother-tongue diversity in nations. American Journal of Sociology
81,1,34-61

\qparr
LIEBERSON (S.) 1985: {Making it count. The improvement of
social research and theory.} University of California Press.

\qparr
LIEBERSON (S.), WATERS (M.C.) 1988: {From many strands. Ethnic 
and racial groups in contemporary America.} Russel Sage. New York.

\qparr
McGARRY (J.), O'LEARY (B.) ed 1993: {The politics of ethnic
conflicts.} Routledge. London. 

\qparr
MACKIE (T.T.), ROSE (R.) 1982: {The international almanac of
electoral history.} Facts on File. New York.

\qparr
MARWELL (G.), OLIVER (P.E.), PRAHL (R.) 1988: Social networks and
collective action. A theory of the critical mass III. American
Journal of Sociology 94,3,502-534.

\qparr
MAY (E.R.) 1973: {``Lessons'' of the past. The use and misuse
of history in American foreign policy.} Oxford University Press.
New York.

\qparr
MILNE (R.S.) 1988: Bicommunal systems: Guyana, Malaysia, Fiji.
Publius: The Journal of Federalism 18,101-113.

\qparr
MOURRE (M.) 1978: {Dictionnaire encyclop\'edique d'histoire.}
Bordas. Paris.

\qparr
OLIVER (P.E.), MARWELL (G.), TEIXEIRA (R.) 1985: A theory of the
critical mass. I: Interdependence, groups, heterogeneity and the
production of collective action. American Journal of Sociology
91,522-556.

\qparr
OLZAK (S.), NAGEL (J.) ed 1986: {Competitive  ethnic
relation.} Academic Press.

\qparr
OLZAK (S.) 1992: {The dynamics of ethnic competition and
conflict.} Stanford University Press. 

\qparr
PASSANT (E.J.) 1960: {A short history of Germany 1815-1945.}
Cambridge University Press. London.

\qparr
PREMDAS (R.R.), SAMARASINGHE (S.W.R. de A.), ANDERSON (A.B.)
eds 1990: {Secessionist movements in comparative perspective.}
Pinter. London.

\qparr
RACINE (J.B.) 1994: Langue et identit\'es territoriales en
Suisse. Annales de G\'eographie 576,152-169.

\qparr
RENNWALD (J.C.) 1984: {La question jurassienne.} Editions
Entente. Paris.

\qparr
RICHARDSON (L.F.)  1960: {Statistics of deadly quarrels.}
Boxwood Press. Pittsburgh.

\qparr
ROBIE (D.) 1989: {Blood on their banner. Nationalist struggles
in the South Pacific.} Zed Books. London.

\qparr
ROEHNER (B.M.) 1993a: Les logiques de l'histoire. Preprint LPTHE
(June)

\qparr
ROEHNER (B.M.) 1993b: A probabilistic reappraisal of Marc Bloch's
comparative approach. Preprint LPTHE (December).

\qparr 
ROEHNER (B.M.) 1997: Spatial and historical determinants
of separatism and integration.
Swiss Journal of Sociology 23,1,25-59.\qL
[This paper is available online in two parts at the following
addresses:\qL
http://www.lpthe.jussieu.fr/$ \sim $roehner/sep1.pdf (part 1)\qL
http://www.lpthe.jussieu.fr/$ \sim $roehner/sep2.pdf (part 2)]

\qparr
ROEHNER (B.M.) 2002: Separatism and integration.
A study in analytical history. Rowman and Littlefield, Laham
(Maryland).

\qparr
ROEHNER (B.M.) 2017: Separatism and disintegration.\qL
[This revised and updated version of the previous publication
is available online at the following address:\qL
http://www.lpthe.jussieu.fr/$ \sim $roehner/separatism.pdf]

\qparr
SHIBUTANI (T.), KWAN (K.M.) 1965: {Ethnic stratification. A
comparative approach.} Macmillan. New York.

\qparr
SIMON (H.) 1959: Theories of decision-making in economics and
behavioral science. The American Economic Review 49,253-283.

\qparr
SNYDER (L.L.) 1976: {Varieties of nationalism: a
comparative study.} Rinehart and Winston. New York. 

\qparr
SNYDER (D.), TILLY (C.) 1972: Harship and collective violence in
France 1830-1960. American Sociological Review 37,771-793.

\qparr
SOROKIN (P.A.) 1937: {Social and cultural dynamics. Vol.III:
Fluctuations and social relationship. War and revolution.}
American Book Company.

\qparr
SVALASTOGA (K.) 1982: Integration, a seven nation comparison.
International Journal of Comparative Sociology 23,3-4,190-202. 

\qparr
THERNSTROM (S.) ed 1980: {Harvard encyclopedia of American
ethnic groups.} Belknap Press of Harvard Unversity Press. Cambridge
(Ma).

\qparr
TILLY (C.): {The contentious French.} The Belknap Press
of Havard University Press. Cambridge,Ma. 

\qparr
TILLY (C.) 1992: How to detect, describe and explain repertories
of contention. New School for Social Research. Working Paper No150.

\qparr
TILLY (C.) 1993: {European revolutions 1492-1992.} Oxford
University Press. Oxford.

\qparr
VERNEIX (J.-C.) 1979: {Les Acadiens.} Editions Entente. Paris. 

\qparr
WEBER (E.) 1976: {Peasants into Frenchmen. The modernization
of rural France 1870-1914.} Stanford University Press. Stanford.

\qparr
WRIGHT (Q.) 1942: {A study of war.} University of Chicago
Press. Chicago.

\qparr
ZAINU'DDIN (A.G.T.) 1968,1980: {A short history of Indonesia.}
The Scarecrow Press. Metuchen (N.J.)

\end{document}